\documentclass[conference]{IEEEtran}
\IEEEoverridecommandlockouts
\usepackage{cite}
\usepackage{amsmath,amssymb,amsfonts}
\usepackage{algorithmic}
\usepackage{graphicx}
\usepackage{textcomp}
\usepackage{hyperref}
\usepackage{subcaption}
\usepackage{tikz}
\usepackage{pifont}
\usepackage{multirow}
\usepackage{color, colortbl}
\usepackage[dvipsnames]{xcolor}
\usepackage{draftwatermark}
\SetWatermarkText{Preprint}   % your text here
\SetWatermarkScale{0.4}           % size
\SetWatermarkColor[gray]{0.9}     % light gray so it doesn’t obscure text
\SetWatermarkAngle{45}            % angle in degrees

\usepackage[a4paper, total={184mm,239mm}]{geometry}
\def\BibTeX{{\rm B\kern-.05em{\sc i\kern-.025em b}\kern-.08em
    T\kern-.1667em\lower.7ex\hbox{E}\kern-.125emX}}
    
\begin{document}

% \title{Congruence Scoring: A Roofline-Inspired Evaluation Of Heterogeneous FPGA Architectures}

\title{Lightweight Congruence Profiling for Early Design Exploration of Heterogeneous FPGAs

\thanks{
This paper has been accepted for presentation at VLSI-SoC in October.\vspace{5mm}
}
}

\newcommand{\TODO}[1]{{\color{red} TODO: #1}}
\newcommand{\BS}[1]{{\color{blue} BS: #1}}

\author{\IEEEauthorblockN{
Allen Boston\IEEEauthorrefmark{1},
Biruk Seyoum\IEEEauthorrefmark{2},
Luca Carloni\IEEEauthorrefmark{2},
Pierre-Emmanuel Gaillardon\IEEEauthorrefmark{1}
}

\IEEEauthorblockA{\IEEEauthorrefmark{1}Department of Electrical and Computer Engineering,
University of Utah, Salt Lake City, UT, 84112 \\
\{allen.boston, pierre-emmanuel.gaillardon\}@utah.edu
}

\IEEEauthorblockA{\IEEEauthorrefmark{2}Department of Computer Science, Columbia University in the City of New York, New York, NY 10027\\
\{biruk, luca\}@cs.columbia.edu
}}

\maketitle

\begin{abstract}
Field-Programmable Gate Arrays (FPGAs) have evolved from uniform logic arrays into heterogeneous fabrics integrating digital signal processors (DSPs), memories, and specialized accelerators to support emerging workloads such as machine learning. While these enhancements improve power, performance, and area (PPA), they complicate design space exploration and application optimization due to complex resource interactions.

To address these challenges, we propose a lightweight profiling methodology inspired by the Roofline model. It introduces three congruence scores that quickly identify bottlenecks related to heterogeneous resources, fabric, and application logic. Evaluated on the Koios and VPR benchmark suites using a Stratix 10–like FPGA, this approach enables efficient FPGA architecture co-design to improve heterogeneous FPGA performance.

\end{abstract}


\section{Introduction}
\label{section:intro}

The development of new Field-Programmable Gate Array (FPGA) architectures typically requires extensive design space exploration (DSE), involving iterative performance evaluation and fine-tuning using a range of benchmark applications~\cite{COFFE}. Over the years, the architecture of FPGAs has undergone significant evolution, transitioning from homogeneous arrays of general-purpose logic to sophisticated fabrics that incorporate heterogeneous compute blocks and memory resources ~\cite{xilinxaiengine, flexlogic_ai, langhammer2021stratix, acrhonix_mlp}. This architectural shift sacrifices generality in favor of higher performance, aiming to narrow the gap between reconfigurable platforms and application-specific integrated circuits (ASICs)~\cite{boutros2021fpga, aman_mat_mult}. Although these enhancements yield substantial gains in power, performance, and area (PPA), they also increase the complexity of balancing application characteristics with architectural resources, adding to the complexity of evaluating candidate FPGA architectures during DSE.

\begin{figure}[t!]
    \centering
    \includegraphics[width=0.95\linewidth]{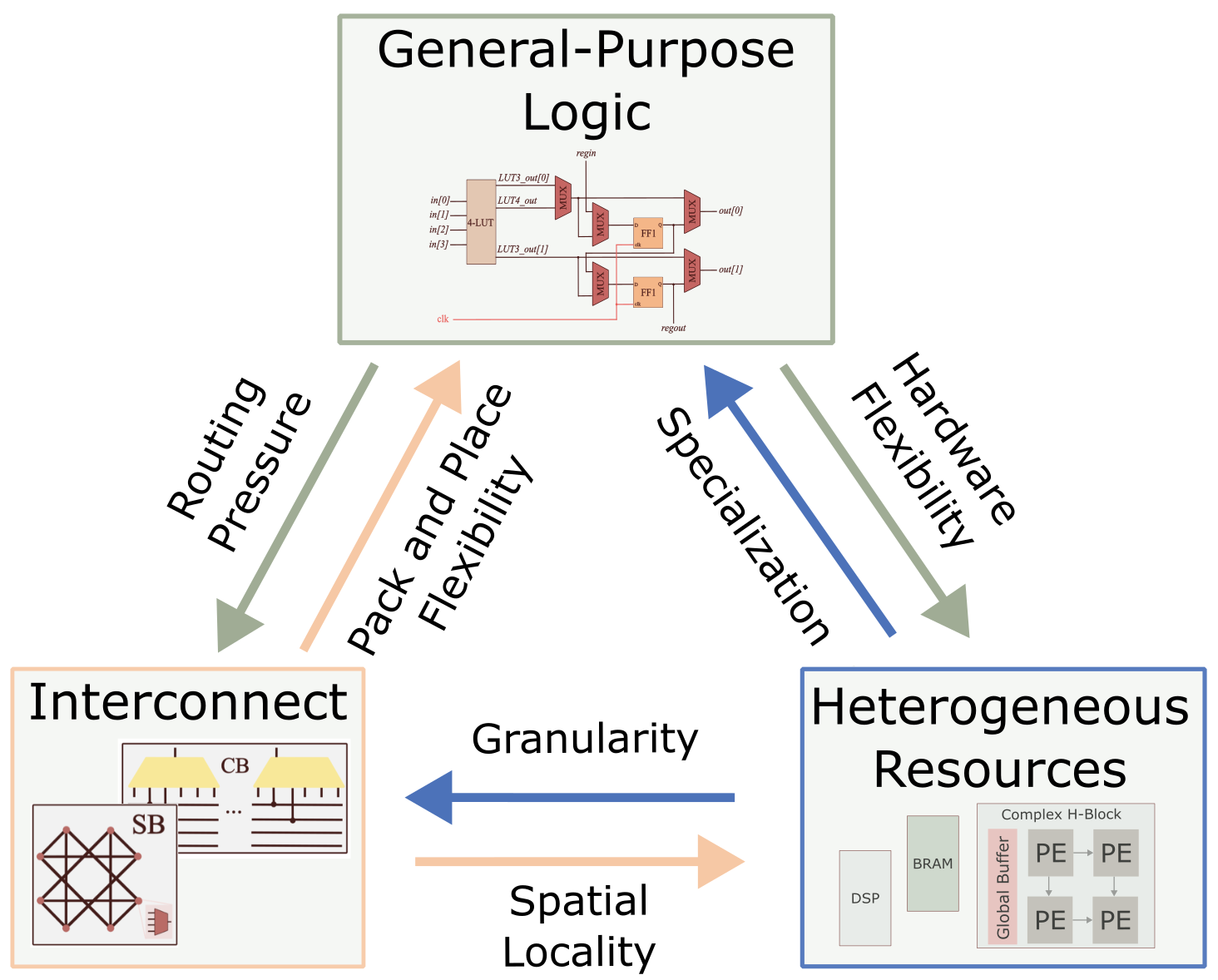}
    \caption{Interplay among logic, heterogeneous resources, and interconnect in FPGA architectures. Arrows show directional influences, while labels indicate pressures such as specialization, flexibility, mapping feasibility that shape congruence between applications and hardware.}
    \label{fig:triangle}
\end{figure}

The performance of traditional FPGA architectures, characterized by limited resource heterogeneity, is largely determined by the complex interplay of spatial layout, resource delay, and the routing locality. As FPGAs evolve to include a broader range of \emph{heterogeneous components} (H-blocks), the interaction between general-purpose logic and specialized resources becomes increasingly critical to overall performance. This intricate relationship is illustrated in Figure~\ref{fig:triangle}, which models the interdependence among three fundamental elements: the general-purpose logic blocks (CLBs), the interconnect, and the specialized resources (H-blocks) embedded within the fabric. Each edge in the triangle represents a direction of influence. The architecture affects both the spatial organization of resources and the class of applications that can be efficiently supported. The application imposes specific demands on specialized blocks such as digital signal processing (DSP) blocks and memory, which may or may not be well-suited for its computation pattern; and finally, the design and distribution of H-blocks within the fabric feeds back into the architecture considerations, introducing placement complexities and increasing routing pressure. Therefore, the increasing heterogeneity of the fabric requires a fundamental rethinking of how FPGA architectures are evaluated and optimized.

\begin{figure*}[ht!]
    \centering
    \includegraphics[width=\linewidth]{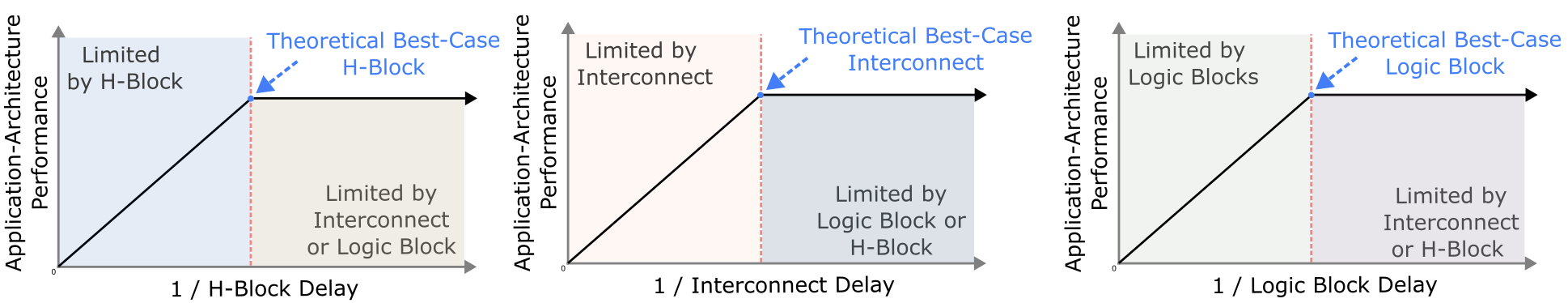}
    \caption{Roofline-inspired model of FPGA subsystem bottleneck contribution. Arrows indicate shifting performance bottlenecks as specific components (e.g., H-blocks or interconnect) are improved, emphasizing the need for balanced architectural optimization. }
    \label{fig:theory}
\end{figure*}

Traditional design space exploration is often manual and ad hoc, overlooking how applications align with FPGA resources or how specialized components affect general-purpose reconfigurability (Figure~\ref{fig:triangle}). Existing CAD tools focus on low-level exploration but provide limited guidance for high-level architectural decisions. As FPGA fabrics grow more complex, fast and interpretable metrics are needed to guide co-design without exhaustive trial-and-error, motivating the development of a lightweight and straightforward profiling methodology.

Building on this insight, we introduce a \emph{congruence score} that quantitatively evaluates how well a heterogeneous FPGA architecture aligns with a given application, enabling efficient and lightweight fabric optimization. Inspired by the Roofline model~\cite{williams2009roofline}, our approach systematically analyzes delays and interconnect characteristics. By focusing on delay, we isolate application–architecture alignment, since specialized compute blocks naturally improve area and power. From this analysis, three congruence scores are derived to highlight key performance bottlenecks in heterogeneous FPGA designs.

\begin{itemize}
\item \textbf{Interconnect Congruence Score (ICS)} — measures how much the physical FPGA fabric limits application performance, revealing whether architectural layout, routing topology, or block placement creates bottlenecks.
\item \textbf{Heterogeneous Resource Congruence Score (HRCS)} — quantifies how well the application utilizes specialized compute blocks, such as DSPs and memory elements, by evaluating their contribution to critical path delays or routing pressure.
\item \textbf{Logic Block Congruence Score (LBCS)} — assesses the alignment between the application’s logic structure and the FPGA’s general-purpose logic resources, indicating how effectively the application utilizes the available fabric.
\end{itemize}

Applying this methodology using a Stratix 10–like FPGA to a subset of the Koios and VPR benchmark suites ~\cite{arora2021koios, VPR}, we demonstrate how these scores assist designers in optimizing FPGA layouts and facilitating faster and more informed co-design decisions. Finally, we pair each evaluated application with its best-fit architecture using the proposed scoring method.

The remainder of this paper is organized as follows. Section \ref{section:method} outlines our proposed scoring methodology. Section \ref{section:exp} demonstrates the application of the congruence scoring system on well-known benchmark suites and presents the results. Section \ref{section:conc} concludes the paper.

% % Designers must evaluate whether an application’s compute pattern aligns with the heterogeneous resources and layout characteristics of the underlying FPGA fabric.
% % Moreover, as reconfigurable platforms become more application-specific, there must be a balance between optimizing for specificity and accommodating a broad range of benchmarks within the same domain. 

\section{Defining Application-Architecture Congruence}
\label{section:method}

In this section, we present details about the congruence score used to assess how effectively an application leverages the various subsystems within a heterogeneous FPGA architecture. Building on a Roofline-inspired model, our approach isolates the contributions of key FPGA subsystems, represents their theoretical performance limits, and quantifies their influence on overall system behavior. 

Figure~\ref{fig:theory} visualizes our conceptual model, showing the interplay between heterogeneous resources, FPGA architecture, and application behavior. For example, when an application relies heavily on heterogeneous resources, improving the timing of those resources initially reduces the critical path. However, beyond a certain point, the critical path shifts to other regions of the fabric, such as routing or general logic, requiring further optimization there. Similarly, applications constrained by routing may benefit from reduced congestion or shorter interconnects up to a limit, after which the logic structure becomes the primary constraint. Finally, if the logic fabric itself dominates timing, further improvements must come from better pipelining, restructuring, or architectural enhancements.

\begin{figure*}[ht!]
    \centering
    \scalebox{0.95}{\includegraphics[width=\linewidth]{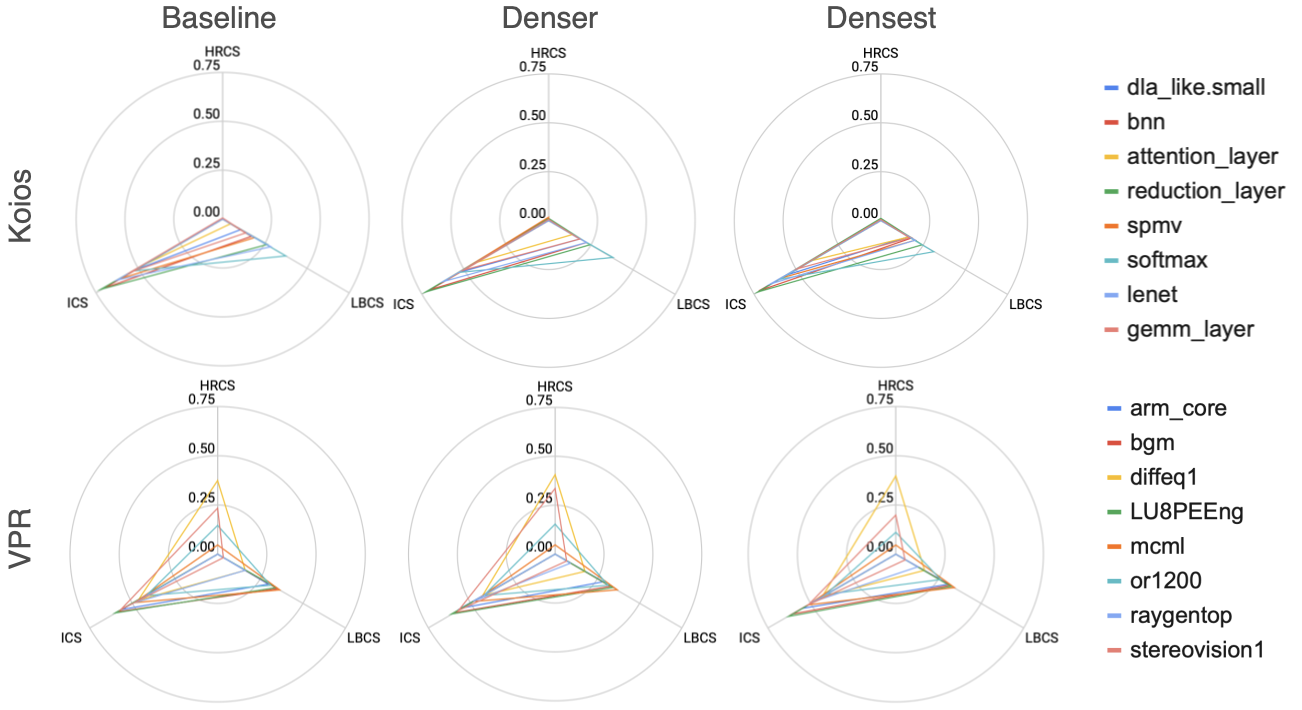}}
    \caption{Radar plots illustrating the congruence scores for a subset of Koios and VPR benchmarks across three architecture variants. The plots highlight how performance bottlenecks shift between heterogeneous resources, interconnect, and logic for varying benchmark sets.}
    \label{fig:results}
\end{figure*}

This evaluation uses VPR~\cite{Titan_VPR}, a versatile architecture modeling tool for exploring reconfigurable platforms. VPR supports diverse configurations, including complex components, custom routing, and heterogeneous compute and memory resources. Its integrated placement and routing engine maps applications onto candidate architectures, enabling accurate performance evaluation. Using this framework, we can modify delays in the architecture description to isolate the theoretical limits of individual FPGA subsystems.
We set these modified delays near-zero to emulate the Roofline ideal for each subsystem. This highlights how each component would contribute to overall performance if unconstrained, revealing the true bottlenecks in the application–architecture mapping.

% \TODO{Here the formulation is generic for all three scores but Luca suggested a better formulation that encompasses all three. For example,  $ \forall i \in \{h\_block, interconnect, logic\}  Score_i = 1 - \frac{\alpha_i - \beta_i}{\gamma_i - \beta_i}$ }
% All three scores share a common formulation that normalizes the change in critical path delay against a user defined ideal performance target:
\vspace{-2mm}
\begin{equation}
\label{eq:score}
 \forall i \in \{ICS, HRCS, LBCS,\} \: Score_i = 1 - \frac{\alpha_i - \beta_i}{\gamma_i - \beta_i} 
\end{equation}

To quantify the application-architecture congruence score, we used Equation \ref{eq:score}. Here, $\alpha_i$ is the delay after modifying one architectural subsystem, $\gamma_i$ is the original timing result with no modifications, and $\beta_i$ is a user-defined target delay. This formulation enables interpretable profiling of how each architectural component contributes to performance limitations. A score approaching zero indicates that the subsystem has minimal impact on the critical path, suggesting limited opportunity for further improvement. In contrast, a score approaching one suggests that the subsystem is a dominant performance bottleneck and a prime target for co-design optimization.

\subsection{Interconnect Congruence Score}

The Interconnect Congruence Score (ICS) measures how much an application's performance is limited by the routing infrastructure. To compute ICS, all interconnect delays are set to near-zero, while logic and heterogeneous block delays remain unchanged. ICS is quantified using Equation~\ref{eq:score}. A high ICS indicates that the routing network significantly contributes to performance degradation. In such cases, architectural strategies like adding direct interconnects or hierarchical routing may help. Similarly, enhancements such as pipelining and buffering can reduce routing delays, although they may require application-level scheduling changes. 

\subsection{Heterogeneous Resource Congruence Score}
The Heterogeneous Resource Congruence Score (HRCS) evaluates how much an application relies on specialized compute and memory elements such as DSPs, BRAMs, or custom accelerators ~\cite{aman_mat_mult, tensor_slice1, comefa}. To compute HRCS, we modify the architecture to assign near-zero delay values to all heterogeneous blocks, simulating ideal performance for these components.

% We then re-run timing analysis and use the resulting delay to assess whether the application remains constrained by other parts of the architecture.

By isolating blocks like DSPs or BRAMs, we can observe whether they form a significant portion of the critical path. If performance improves substantially under idealized heterogeneous block conditions, it indicates that the application is dependent on those resources and that they may be performance bottlenecks. While we group all specialized blocks into a single HRCS value for simplicity, the methodology can be extended to separately evaluate each component type. For example, one could isolate BRAMs or DSPs individually to determine their relative contributions to overall performance limitations.

%By isolating the DSPs, we can directly observe the theoretical performance limit illustrated in Figure~\ref{fig:theory}. Identifying this limit provides insight into how dependent the application is on DSP resources and whether they represent a performance bottleneck that warrants further optimization. While this work maintains a high-level abstraction for simplicity by merging all heterogeneous blocks into a single score, the methodology can be extended to evaluate heterogeneous resources individually. For example, the approach could be used to separately quantify and compare the performance impact of BRAMs versus DSPs within the same architectural context.

\subsection{Logic Block Congruence Score}

The Logic Block Congruence Score (LBCS) quantifies the extent to which an application's internal logic structure constrains performance. Many FPGA applications achieve performance gains by using deep pipelining, inserting memory elements between stages of combinational logic to shorten the critical path. However, if the underlying logic fabric remains a bottleneck, further optimization may be needed at the application or architectural level.

To compute LBCS, the delays of all logic elements such as LUTs and multiplexers are set to zero, while delays for routing and heterogeneous resources remain unchanged. A high LBCS means logic is the dominant limiter and may require architectural changes (e.g., larger LUTs or more local interconnect) or software-level improvements like better pipelining and scheduling. A low LBCS, on the other hand, suggests that the logic fabric is not a critical contributor to performance constraints.

\section{Experimental Setup and Results}
\label{section:exp}

\subsection{Benchmark Selection and Target Architectures}

We evaluate our profiling methodology using a subset of benchmarks from the Koios and VPR suites \cite{arora2021koios, koios2.0, VPR}, including compute-intensive machine learning tasks such as an Intel DLA–like accelerator and convolution layers, as well as general-purpose compute designs from VPR like an ARM core. The benchmarks target an FPGA architecture provided by the Koios framework, modeled after a Stratix 10 routing structure, with complex DSPs inspired by Agilex, configurable 20Kb BRAMs, and CLBs composed of 10 6-input LUTs.

To examine the impact of heterogeneous resource density, we study three architecture variants. The first is a \emph{baseline} with a balanced resource mix similar to commercial FPGAs. The second, \emph{denser}, increases the number of DSPs and BRAMs. The third, \emph{densest}, further increases the ratio of specialized blocks to general-purpose logic. These variants form the foundation of our analysis.

This work builds on the original Koios study by extending the evaluation beyond traditional metrics like wire length and critical path delay \cite{koios2.0}. Instead, we focus on the architectural interplay among general-purpose logic, routing fabric, and heterogeneous blocks, providing a multidimensional view of how well architectures align with application demands.

%All architectures are annotated with timing values based on a 22 nanometer technology node. These delay annotations were generated using COFFE.

\subsection{Experimental Procedure}

For each architecture variant, we compute HRCS, ICS, and LBCS across selected Koios and VPR benchmarks using the VPR flow. This is achieved by modifying the architecture description files to isolate subsystem delays, as described in Section~\ref{section:method}, so that each score reflects the impact of a specific architectural component while preserving realistic placement and routing behavior. We first run VPR with full delay annotations to obtain baseline packing, placement, routing, and critical path delays. The resulting packing, placement, and routing information are then reused in subsequent runs with zeroed subsystem delays to prevent VPR from executing the full flow again. In these runs, only the final timing analysis is performed to extract the subsystem-specific impact.
% The resulting weighted score enables a quantitative and interpretable evaluation of overall application–architecture alignment.

\subsection{Results}

\begin{table}
    \centering
    \begin{tabular}{|c|c|c|c|} \hline
       \cellcolor{Gray!5} \textbf{Koios Benchmarks} & \cellcolor{Gray!5}\textbf{Baseline} & \cellcolor{Gray!5}\textbf{Denser} & \cellcolor{Gray!5}\textbf{Densest}   \\ \hline
        \cellcolor{Gray!5} dla\_like.small          & \cellcolor{ForestGreen!50} 0.545  & \cellcolor{Yellow!50} 0.565 & \cellcolor{Red!50} 0.573   \\ \hline
        \cellcolor{Gray!5} bnn                      & \cellcolor{ForestGreen!50} 0.722   & \cellcolor{Yellow!50} 0.737  & \cellcolor{Red!50} 0.740 \\ \hline
        \cellcolor{Gray!5} attention\_layer         & \cellcolor{Red!50} 0.485 & \cellcolor{Yellow!50} 0.462 & \cellcolor{ForestGreen!50} 0.448    \\ \hline
        \cellcolor{Gray!5} reduction\_layer         & \cellcolor{ForestGreen!50} 0.765 & \cellcolor{Yellow!50} 0.766   & \cellcolor{Red!50} 0.766 \\ \hline
        \cellcolor{Gray!5} spmv                     & \cellcolor{Red!50} 0.643 & \cellcolor{ForestGreen!50}  0.600  & \cellcolor{Yellow!50} 0.626 \\ \hline
        \cellcolor{Gray!5} softmax                  & \cellcolor{ForestGreen!50} 0.644 & \cellcolor{Yellow!50} 0.653   & \cellcolor{Red!50} 0.678 \\ \hline
        \cellcolor{Gray!5} lenet                    & \cellcolor{Red!50} 0.684 & \cellcolor{ForestGreen!50} 0.668   & \cellcolor{Yellow!50} 0.682 \\ \hline
        \cellcolor{Gray!5} gemm\_layer              & \cellcolor{Yellow!50} 0.562 & \cellcolor{Red!50}  0.534  & \cellcolor{ForestGreen!50} 0.522 \\ \hline \hline
        \cellcolor{Gray!5} Koios Mean               & \cellcolor{Yellow!50} 0.561 & \cellcolor{ForestGreen!50} 0.554   & \cellcolor{Red!50} 0.600 \\ \hline \hline
        
        \cellcolor{Gray!5} \textbf{VPR Benchmarks}  & \cellcolor{Gray!5}\textbf{Baseline} & \cellcolor{Gray!5}\textbf{Denser} & \cellcolor{Gray!5}\textbf{Densest}   \\ \hline
        \cellcolor{Gray!5} arm\_core                & \cellcolor{Red!50} 0.636  & \cellcolor{ForestGreen!50} 0.626 & \cellcolor{Yellow!50}  0.630 \\ \hline
        \cellcolor{Gray!5} bgm                      & \cellcolor{Red!50} 0.686  & \cellcolor{ForestGreen!50} 0.684 & \cellcolor{Yellow!50}  0.685 \\ \hline
        \cellcolor{Gray!5} diffeq1                  & \cellcolor{Yellow!50} 0.621 & \cellcolor{Red!50} 0.614   & \cellcolor{ForestGreen!50} 0.613 \\ \hline
        \cellcolor{Gray!5} LU8PEEng                 & \cellcolor{ForestGreen!50} 0.692 & \cellcolor{Yellow!50} 0.700   & \cellcolor{Red!50} 0.706 \\ \hline
        \cellcolor{Gray!5} mcml                     & \cellcolor{Yellow!50} 0.608 & \cellcolor{ForestGreen!50}  0.606  & \cellcolor{Red!50} 0.618 \\ \hline
        \cellcolor{Gray!5} or1200                   & \cellcolor{Yellow!50} 0.543 & \cellcolor{Red!50}  0.545  & \cellcolor{ForestGreen!50} 0.499 \\ \hline
        \cellcolor{Gray!5} raygentop                & \cellcolor{Red!50} 0.512 & \cellcolor{ForestGreen!50}  0.493  & \cellcolor{Yellow!50} 0.497 \\ \hline
        \cellcolor{Gray!5} stereovision1            & \cellcolor{Yellow!50} 0.612 & \cellcolor{Red!50}  0.665  & \cellcolor{ForestGreen!50} 0.555 \\ \hline \hline
        \cellcolor{Gray!5} VPR Mean                 & \cellcolor{Yellow!50} 0.614 & \cellcolor{Red!50}  0.617  & \cellcolor{ForestGreen!50} 0.601 \\ \hline \hline
       
       \cellcolor{Gray!5} \textbf{VPR/Koios Aggregate} & \cellcolor{Red!50} 1.175 & \cellcolor{Yellow!50}  1.170  & \cellcolor{ForestGreen!50} 1.160 \\ \hline
        
    \end{tabular}
    \caption{Application–architecture congruence scores for each benchmark across three architecture variants. The scores highlight best-fit architectures for individual applications and overall trends across the Koios and VPR benchmark sets.}
    \label{tab:table_1}
\end{table}

Figure~\ref{fig:results} shows the results of our architectural exploration as radar plots, where each benchmark from a subset of Koios and VPR is evaluated using the three congruence scores from Section~\ref{section:method}, computed with Equation~\ref{eq:score} using an optimistic ideal delay of 0.2 ns. As shown by the radar plots, Koios applications are primarily limited by routing, followed by logic, with minimal dependence on heterogeneous blocks. This suggests that the H-blocks are well-aligned with the application’s needs, effectively supporting computation without becoming bottlenecks, as evidenced by the fact that critical paths seldom pass through them \cite{ganesh_hblock}. As H-block density increases, we observe variations in the influence of general-purpose logic, while routing consistently remains the dominant limiting factor. In contrast, the VPR benchmarks exhibit a more balanced distribution of bottlenecks across the evaluated resources, with greater reliance on heterogeneous components when compared to Koios benchmarks, indicating that their critical paths often include H-blocks. Additionally, the smaller radar plot areas for Koios benchmarks suggest that they are likely to perform better on this architecture set compared to the VPR designs.

While the radar plots provide intuitive visual insight into the independent contributions of heterogeneous and general-purpose resources, we introduce an aggregate scoring methodology to quantitatively assess overall application–architecture alignment. The application–architecture congruence score is calculated as the magnitude of an n-D vector (HRCS, LBCS, ICS), which can be extended to additional dimensions. These scores are summarized in Table~\ref{tab:table_1}. Although the best-fit architecture varies across individual applications, averaging across each benchmark set reveals clear trends: the \emph{Denser} architecture best matches Koios benchmarks, while the \emph{Densest} architecture aligns best with VPR designs. When both benchmark sets are combined, the aggregate score indicates that the \emph{Densest} architecture provides the best overall fit. This is likely due to the strong compatibility of Koios benchmarks with H-block–rich architectures and the relatively minor performance shifts in VPR designs as H-block density increases.

\section{Conclusion}
\label{section:conc}

We present a lightweight profiling methodology that quantifies the alignment between applications and heterogeneous FPGA architectures using congruence scores. These scores isolate bottlenecks in specialized  compute blocks, routing, and general-purpose logic based on the roofline model. Applied to well adopted benchmark suites and FPGA architectures, our method quickly identifies architectural constraints and supports effective co-design decisions. It enables faster design space exploration and can be integrated into automated optimization tools, thereby advancing the understanding of application–architecture interactions in FPGAs.

\bibliographystyle{IEEEtran}
\bibliography{IEEEabrv,main}

% Generated by IEEEtran.bst, version: 1.14 (2015/08/26)
\begin{thebibliography}{10}
\providecommand{\url}[1]{#1}
\csname url@samestyle\endcsname
\providecommand{\newblock}{\relax}
\providecommand{\bibinfo}[2]{#2}
\providecommand{\BIBentrySTDinterwordspacing}{\spaceskip=0pt\relax}
\providecommand{\BIBentryALTinterwordstretchfactor}{4}
\providecommand{\BIBentryALTinterwordspacing}{\spaceskip=\fontdimen2\font plus
\BIBentryALTinterwordstretchfactor\fontdimen3\font minus \fontdimen4\font\relax}
\providecommand{\BIBforeignlanguage}[2]{{%
\expandafter\ifx\csname l@#1\endcsname\relax
\typeout{** WARNING: IEEEtran.bst: No hyphenation pattern has been}%
\typeout{** loaded for the language `#1'. Using the pattern for}%
\typeout{** the default language instead.}%
\else
\language=\csname l@#1\endcsname
\fi
#2}}
\providecommand{\BIBdecl}{\relax}
\BIBdecl

\bibitem{COFFE}
\BIBentryALTinterwordspacing
S.~Yazdanshenas and V.~Betz, ``Coffe 2: Automatic modelling and optimization of complex and heterogeneous fpga architectures,'' \emph{ACM Trans. Reconfigurable Technol. Syst.}, vol.~12, no.~1, jan 2019. [Online]. Available: \url{https://doi.org/10.1145/3301298}
\BIBentrySTDinterwordspacing

\bibitem{xilinxaiengine}
``Versal ai engine,'' \url{https://www.xilinx.com/products/intellectual-property/versal-ai-engine.html#overview}, 2023.

\bibitem{flexlogic_ai}
``Flexlogix inferx ai,'' \url{https://flex-logix.com/inferx-ai/inferx-ai-hardware/}, 2023.

\bibitem{langhammer2021stratix}
M.~Langhammer, E.~Nurvitadhi, B.~Pasca, and S.~Gribok, ``Stratix 10 nx architecture and applications,'' in \emph{The 2021 ACM/SIGDA International Symposium on Field-Programmable Gate Arrays}, 2021, pp. 57--67.

\bibitem{acrhonix_mlp}
``Achronix machine learning processor,'' \url{https://www.achronix.com/machine-learning-processor}, 2023.

\bibitem{boutros2021fpga}
A.~Boutros and V.~Betz, ``Fpga architecture: Principles and progression,'' \emph{IEEE Circuits and Systems Magazine}, vol.~21, no.~2, pp. 4--29, 2021.

\bibitem{aman_mat_mult}
\BIBentryALTinterwordspacing
A.~Arora, Z.~Wei, and L.~John, ``The case for hard matrix multiplier blocks in an fpga,'' in \emph{Proceedings of the 2020 ACM/SIGDA International Symposium on Field-Programmable Gate Arrays}, ser. FPGA '20.\hskip 1em plus 0.5em minus 0.4em\relax New York, NY, USA: Association for Computing Machinery, 2020, p. 323. [Online]. Available: \url{https://doi.org/10.1145/3373087.3375360}
\BIBentrySTDinterwordspacing

\bibitem{williams2009roofline}
S.~Williams, A.~Waterman, and D.~Patterson, ``Roofline: an insightful visual performance model for multicore architectures,'' \emph{Communications of the ACM}, vol.~52, no.~4, pp. 65--76, 2009.

\bibitem{arora2021koios}
A.~Arora, A.~Boutros, D.~Rauch, A.~Rajen, A.~Borda, S.~A. Damghani, S.~Mehta, S.~Kate, P.~Patel, K.~B. Kent \emph{et~al.}, ``Koios: A deep learning benchmark suite for fpga architecture and cad research,'' in \emph{2021 31st International Conference on Field-Programmable Logic and Applications (FPL)}.\hskip 1em plus 0.5em minus 0.4em\relax IEEE, 2021, pp. 355--362.

\bibitem{VPR}
V.~Betz and J.~Rose, ``Vpr: A new packing, placement and routing tool for fpga research,'' in \emph{Proceedings of the 7th International Workshop on Field-Programmable Logic and Applications}, ser. FPL '97.\hskip 1em plus 0.5em minus 0.4em\relax Berlin, Heidelberg: Springer-Verlag, 1997, p. 213–222.

\bibitem{Titan_VPR}
\BIBentryALTinterwordspacing
K.~E. Murray, S.~Whitty, S.~Liu, J.~Luu, and V.~Betz, ``Timing-driven titan: Enabling large benchmarks and exploring the gap between academic and commercial cad,'' \emph{ACM Trans. Reconfigurable Technol. Syst.}, vol.~8, no.~2, Mar. 2015. [Online]. Available: \url{https://doi.org/10.1145/2629579}
\BIBentrySTDinterwordspacing

\bibitem{tensor_slice1}
\BIBentryALTinterwordspacing
A.~Arora, S.~Mehta, V.~Betz, and L.~K. John, ``Tensor slices to the rescue: Supercharging ml acceleration on fpgas,'' in \emph{The 2021 ACM/SIGDA International Symposium on Field-Programmable Gate Arrays}, ser. FPGA '21.\hskip 1em plus 0.5em minus 0.4em\relax New York, NY, USA: Association for Computing Machinery, 2021, p. 23–33. [Online]. Available: \url{https://doi.org/10.1145/3431920.3439282}
\BIBentrySTDinterwordspacing

\bibitem{comefa}
\BIBentryALTinterwordspacing
A.~Arora, A.~Bhamburkar, A.~Borda, T.~Anand, R.~Sehgal, B.~Hanindhito, P.-E. Gaillardon, J.~Kulkarni, and L.~K. John, ``Comefa: Deploying compute-in-memory on fpgas for deep learning acceleration,'' \emph{ACM Trans. Reconfigurable Technol. Syst.}, vol.~16, no.~3, jul 2023. [Online]. Available: \url{https://doi.org/10.1145/3603504}
\BIBentrySTDinterwordspacing

\bibitem{koios2.0}
A.~Arora, A.~Boutros, S.~A. Damghani, K.~Mathur, V.~Mohanty, T.~Anand, M.~A. Elgammal, K.~B. Kent, V.~Betz, and L.~K. John, ``Koios 2.0: Open-source deep learning benchmarks for fpga architecture and cad research,'' \emph{IEEE Transactions on Computer-Aided Design of Integrated Circuits and Systems}, vol.~42, no.~11, pp. 3895--3909, 2023.

\bibitem{ganesh_hblock}
A.~Mishra, N.~Rao, G.~Gore, and X.~Tang, ``Architectural exploration of heterogeneous fpgas for performance enhancement of ml benchmarks,'' in \emph{2023 IEEE Asia Pacific Conference on Circuits and Systems (APCCAS)}, 2023, pp. 232--235.

\end{thebibliography}

\end{document}